# Absence of zero-energy surface bound states in $Cu_xBi_2Se_3$ via a study of Andreev reflection spectroscopy

Haibing Peng *, Debtanu De, Bing Lv, Fengyan Wei, Ching-Wu Chu [†]

Department of Physics and the Texas Center for Superconductivity, University of Houston,

Houston, Texas 77204-5005, USA

* haibingpeng@uh.edu

[†] cwchu@uh.edu

ABSTRACT

$Cu_xBi_2Se_3$ has been proposed as a potential topological superconductor characterized by an odd-parity full bulk superconducting gap and zero-energy surface Andreev bound states (Majorana fermions). A consequence of such Majorana fermions is a peak in the zero-energy density of states which should lead to a persistent zero-bias-conductance-peak (ZBCP) in Andreev reflection (AR) or tunneling experiments. Here we employ a newly developed nanoscale AR spectroscopy method to study normal metal/superconductor (N-S) devices featuring $Cu_xBi_2Se_3$. The results show that a ZBCP can be tuned in or out from $Cu_xBi_2Se_3$ samples depending on the N-S barrier strength. While the appearance of ZBCP may be traced to different origins, its absence under finite barrier strength represents the absence of zero-energy Majorana fermions. The present observations thus call for a reexamination of the nature of the superconducting state in $Cu_xBi_2Se_3$.

Novel quantum states of matter with topological characteristics have been emerging as exciting research topics since the discovery of topological insulators.[1-5] Protected by time-reversal symmetry, a topological insulator (e.g., $Bi_2Se_3$) [6, 7] possesses a full band gap in the bulk, but has gapless surface states with an odd number of Dirac points (Kramers degenerate band crossing). In analogy, with particle-hole symmetry considered, topological superconductors [8-14] were theoretically predicted to exist with a full bulk superconducting gap but stable gapless Majorana surface bound states. Under active research currently is the experimental search for topological superconductors and initial potential candidates include the $^3$He-B phase [8, 9]and the interface between a topological insulator and an s-wave superconductor.[10] Recently, superconductivity below Tc ~ 3.8 K was observed [15]in an electron-doped topological insulator $Cu_xBi_2Se_3$, and a subsequent theory[16] has proposed that $Cu_xBi_2Se_3$ could be a potential candidate of topological superconductors using a phenomenological model predicting the existence of an odd-parity full superconducting gap with time-reversal symmetry and the resultant zero-energy surface Andreev bound states (Majorana fermions). This appears to be confirmed [17] by the successful detection of a zero-bias-conductance-peak (ZBCP) as the evidence for the existence of zero-energy Majorana fermions by a "soft" point-contact technique. In view of the significance of the observation and the complexity involved in point-contact devices, further stringent experimental test on the existence of zero-energy Majorana fermions is called for. Here we have employed a newly developed nanoscale Andreev reflection (AR) spectroscopy method [18] to elucidate the superconducting gap structures of $Cu_xBi_2Se_3$ as a function of temperature and magnetic field. The results show that the ZBCP can be tuned in or out from $Cu_xBi_2Se_3$ samples depending on the normal metal/superconductor (N-S) barrier strength Z. While the appearance of ZBCP may be traced to different origins, its absence under finite barrier strength represents the absence of zero-energy Majorana fermions. The present observations thus call for a reexamination of the nature of the superconducting state in $Cu_xBi_2Se_3$, presumably a representative (if not the only) topological superconductor discovered to date.

The discovery of superconductivity in $Cu_xBi_2Se_3$ [15] has generated great interest since the parent compound $Bi_2Se_3$ is a time-reversal-invariant topological insulator, and immediately drew attention [16, 19] on its implication for the search of Majorana fermions and topological superconductors. It has been suggested theoretically that an odd parity pairing could be induced due to strong spin-orbital coupling in $Cu_xBi_2Se_3$, which should then meet the criteria for a topological superconductor since it is a time-reversal-invariant centrosymmetric superconductor with a full odd-parity pairing gap and a Fermi surface enclosing an odd number of time-reversal-invariant momentum in the Brillouin zone.[16] In this model, two-orbital effective short-range electron interactions are assumed to be responsible for the superconductivity and as a result a topological superconductor phase occurs for certain range of interaction strength. Majorana fermions then manifest themselves in a Kramers pair of zero-energy surface Andreev bound states. Therefore, a central idea for experimentally identifying the topological superconductor phase is to detect a peak of density of states at zero energy. Recently, a ZBCP was observed in an Andreev reflection spectroscopy study of $Cu_xBi_2Se_3$ in N-S junctions made from a soft point-contact method,[17] and this result was interpreted as the evidence for topological superconductivity in $Cu_xBi_2Se_3$. However, if there does exist a peak in the zero-energy density of states (as a result of the existence of Majorana fermions), the ZBCP should occur persistently in all N-S junctions with different barrier strength $Z$ ranging from the transparent limit to the tunneling limit. Unfortunately, the relatively low Tc (low energy scale) and possible electronic phase separation in $Cu_xBi_2Se_3$ [15] post enormous challenges for traditional point-contact AR experiments. In this work, we employ a unique experimental method to construct nano-scale devices of N-S junctions featuring $Cu_xBi_2Se_3$ with different barrier strength $Z$ and study the Andreev reflection spectroscopy as a function of magnetic field and temperature for N-S junctions with different $Z$ parameters. Our results show that the ZBCP appears in N-S junctions with transparent barriers, but disappears in N-S junctions with finite barriers showing archetypical double-peak AR spectra. The occurrence of a ZBCP in $Cu_xBi_2Se_3$ can be consistently explained as the Andreev reflection process in combination with the effect of the critical current and thus rules out the existence of zero-energy surface bound states in $Cu_xBi_2Se_3$.

We started with single-crystal bulk materials of $Cu_xBi_2Se_3$ with $x \sim 0.15$, similar to previous reports. [15] Single crystals of $Cu_xBi_2Se_3$ up to cm size with c-axis preferred orientation are grown by slowly cooling of the melting mixture as described in Ref. [15]. The reactant mixture of Bi(99.999% pure), Cu (99.995% pure), and Se (99.999% pure), is sealed in a quartz tube under vacuum, heated up to 860 $^oC$ for 20 hrs, slowly cooled down to 600 $^oC$ over 3 days, and then quenched in cold water. X-ray diffraction pattern in θ-2θ scan confirms the single crystallinity of the samples and measured rocking curves show a full width at half maximum of 0.3 degree for the (0 0 15) peak. Magnetic susceptibility measurement (Fig. 1a) indicates an on-set bulk Tc ~3.4 K and a superconducting volume fraction up to ~ 20 % at 2 K, close to results previously reported.[15]

We have employed a newly developed technique [18] to construct nanoscale N-S devices and perform Andreev reflection spectroscopy. In brief, we mechanically cleaved the bulk material of $Cu_xBi_2Se_3$ into micro-scale crystals, immediately transferred them into a vacuum chamber (~ $10^{-5}$ Torr), and used a sharp probe tip attached on a micro-manipulator to place a piece of microcrystal on top of multiple parallel metal electrodes spaced ~ 500 nm apart as designed via electron-beam lithography (Fig. 1b). To obtain AR spectra for a target N-S junction, we used a special circuit [18] as illustrated in Fig. 1c where a small AC current superimposed to a DC bias current is applied between the $I_+$ and $I_-$ terminals while both the DC and the AC voltages across the N-S junction are measured between the $V_+$ and $V_-$ terminals. Therefore, the AR spectrum, i.e. the differential conductance $dI/dV$ vs. the DC voltage $V$ across the target junction between the $I_+$ ($V_+$) terminal and the superconductor, is obtained. The flexibility in selecting the configuration for aforementioned measurement terminals enables us to study multiple N-S junctions between different metal electrodes and the same superconductor crystal with different $Z$ parameters in a single device.

Fig. 2 shows the normalized $dI/dV$ vs. $V$ at $T = 240$ mK for three N-S junctions. Two N-S junctions are from the device in Fig. 1b and the third one is from another device. For all three junctions of Figs. 2a-2c, the AR spectra (symbols) demonstrate a $dI/dV$ dip at zero bias and two shoulders at $V \sim \pm 0.4$ mV, which is typical for N-S interface with finite barrier strength $Z$ according to the BTK theory.[20] The solid lines in Fig. 2 represents a fitting of the experimental $dI/dV$ data (normalized to the normal state data above Tc) by the generalized BTK theory considering the broadening effect.[21, 22] Considering the coexistence of superconducting and non-superconducting phases in bulk $Cu_xBi_2Se_3$, [15] we can express the total normalized conductance as $\sigma = w\sigma_s + (1-w)$, where $\sigma_s$ is the normalized conductance for the interface between the normal metal and the superconducting phase, calculated according to the BTK theory,[21] and $w$ is the weight of contribution to conductance from the superconducting phase in the nano-scale junction ($w$ is related to the superconducting volume fraction). From the fitting of the three AR spectra in Fig. 2, we obtain the energy of the SC gap as $\Delta = 0.35 \pm 0.04$ meV (with the broadening parameter $\Gamma$ ranging from 0.13 to 0.27 meV). The weight of contribution for the superconducting phase varies significantly from $w = 13\%$ to 80%, which indicates electronic phase separation on microscopic scale and is consistent with the low superconducting volume fraction (up to ~ 20%) as measured for the bulk material. Notably, despite of different barrier strength and junction characteristics, all three N-S junctions from different samples reproducibly give similar values of a superconducting energy gap $\Delta \sim 0.35$ meV, demonstrating that spectroscopic, energy-resolved information of the superconducting gap is achieved. We note that those spectroscopic N-S junctions (Fig. 2) usually show a large normal state resistance $R_N$ (> 100 $\Omega$), but for junctions with $R_N$ smaller than 10 $\Omega$ (Fig. 3) the energy scale is not accurate and we usually observe a ZBCP followed by a dip at the peak edge which we will discuss in detail later. Our observation of spectroscopic junctions is consistent with a commonly adopted empirical rule [23] for determining ballistic transport regime: $R_N \gg 4\rho/3\pi l$, with $l$ the mean free path of electrons in the junction and $\rho$ the bulk material resistivity. This empirical rule is derived considering that the point contact size has to be less than the mean free path to ensure ballistic transport across the N-S interface which is sufficient for providing spectroscopic, energy-resolved information of the superconducting gap (see Ref. [24] for more discussions). By this rule, we can reach a conclusion of selecting $R_N$ much larger than ~13 $\Omega$ as the criteria for ballistic N-S point contacts in $Cu_xBi_2Se_3$ by using bulk $\rho \sim 140$ $\mu\Omega$cm and $l \sim 45$ nm.[25]

In contrast, the junctions with smaller $R_N$ (Fig. 3) display a conductance peak at zero bias with dips at its edge positions, which is typical for nearly transparent N-S interface with weak barrier strength. The ZBCP can in principle have various potential sources of origin, such as the Andreev reflection plateau resulting from bulk superconducting gaps, or a zero-energy density-of-states peak due to potential surface Andreev bound states.[17] However, the absence of a zero-bias conductance peak in the junctions of Fig. 2 with finite barrier strength $Z$ indicates that the ZBCP observed in $Cu_xBi_2Se_3$ is unlikely due to the existence of zero-energy surface Andreev bound states. Instead, the zero bias peak can be explained as the Andreev reflection plateau due to the bulk superconducting gap but with a cut-off of plateau width when the critical current is reached at relatively low bias $V < \Delta/e$ because of the low junction resistance.[18] For the junctions of Figs. 3b and 3c, the width of the zero-bias peak does not reflect the gap energy correctly, because the normal state resistance $R_N$ is less than the aforementioned threshold $R_N$ value for ballisticity (~ 13 Ω) and thus these junctions are not in the regime of ballistic transport due to an estimated point contact size larger than the electron mean free path. On the other hand, for the junction of Fig. 3a, $R_N$ is close to the threshold $R_N$ value of ~ 13 Ω, and thus the peak width is similar to the energy gap ~ 0.35 meV as obtained from Fig. 2. To better elucidate the nature of the ZBCP, we performed new measurements after a period of six months for these two low-conductance junctions of Figs. 3b and 3c. The new results (see Fig. S2 of Ref. [24]) show that their new normal-state resistance $R_N$ is increased to ~ 100 Ω (an order of magnitude higher than the initial values in Figs. 3b and 3c) and the extracted gap values are now similar to those from the junctions of Fig. 2, which is consistent with our previous discussion on the determination of spectroscopic conditions based on $R_N$ values and confirms that the observed ZBCP should be a consequence of the bulk superconducting gap.

In Fig. 4, we present the evolution of the Andreev reflection spectra as a function of both magnetic field and temperature for one N-S junction with large $R_N$ (finite barrier strength) and another with smaller $R_N$ (transparent barrier).[24] For the junction with finite barrier strength (Fig. 4a), as magnetic field is increased, the double peaks (characterizing the superconducting gap) in the AR spectrum become less and less pronounced up to B = 3T, and the double-peak feature completely disappear at B = 5 T which is close to the reported upper critical-field $B_{c2}$ for bulk $Cu_xBi_2Se_3$.[25] We note that the double dips at high bias move towards the center in a much faster pace than that for the double peaks, because the double dips are associated with the breaking of the superconducting state after the critical current is reached and their position change reflects a decreasing critical current as the B field (or temperature) increases.[18] As a function of temperature (Fig. 4b), the double peaks (or superconducting gap) shrinks slowly up to T = 1.1 K. Above T = 1.5 K, the Andreev reflection spectrum changes into a single broad peak (as often observed for other finite-barrier junctions [24]), which can be understood within the BTK theory as the reduction of barrier strength $Z$ due to thermally activated transport across the interface. The *dI/dV* spectrum becomes V-shaped above T = 2 K but continues to evolve even at T = 4.3 K which is above bulk Tc.

For the junction with transparent barrier (Fig. 4c), the zero-bias conductance peak is greatly suppressed at B = 3 T and already disappears at B = 5 T. As a function of temperature (Fig. 4d), a broad bump persists above bulk Tc and disappears at a temperature T ~ 4.5 K (for which the mechanism is not clear). Notable features besides the zero-bias peak are the *dI/dV* oscillations outside of the superconducting gap but before the reach of the normal state. With the increase of temperature, the position of such oscillations shifts towards lower bias, until they disappear at T = 4K. These oscillations are occasionally observed in N-S junctions with small $R_N$ (low barrier) as seen in Fig. 3. As discussed before, the broad dip outside of the gap can be attributed to the role of the critical current, although the origin of the observed *dI/dV* oscillations within the broad dip is not clear. Nevertheless, the overall evolution of the AR spectra as a function of magnetic field and temperature can be reasonably interpreted as a result for a bulk-gap superconductor.[20, 26]

In Ref. [17], a zero-energy density-of-states peak due to Majorana fermions was invoked to explain a ZBCP observed therein. However, we note in particular that the representative AR spectra in Ref. [17] show a ZBCP on top of a strong non-saturating V-shape *dI/dV* background, and the ZBCP observed therein disappears completely in a magnetic field ~ 0.45 T (which is an order of magnitude smaller than the reported upper critical-field value for bulk $Cu_xBi_2Se_3$ [25]). To address this issue, we have performed control experiments to produce N-S junctions showing either a ZBCP with a strong non-saturating V-shape background (Figs. 5a - 5c) or a pure V-shape *dI/dV* spectrum without ZBCP (Figs. 5d – 5f). It is important to note that these two N-S junctions of Fig. 5 were not electrically connected initially after placing a superconductor microcrystal on top of parallel metal electrodes, but electrical connection was later established after a short voltage pulse (~ 10 V in magnitude) was applied to anneal the contacts between the electrodes and the superconductor microcrystal. As shown in Fig. 5, these two N-S junctions (connected after voltage annealing) show a common strong non-saturating V-shape background, with one having a ZBCP. The ZBCP (Fig. 5b) disappears at a magnetic field only ~ 0.4T (similar to what was reported in Ref. [17]), indicating a thick degraded superconducting surface layer dominating the N-S transport. As a comparison, the other annealed junction (Figs. 5d - 5f) shows a pure strong V-shape spectrum with no hint for AR signature, which can be explained by a thick non-superconducting surface layer (we also note that the V-shape background for both junctions of Fig. 5 shows similar temperature and B-field dependence which cautions any interpretation by the so-called "pseudo-gap" behavior above Tc [17]). In stark contrast, for two additional junctions in the same device of Fig. 5 which were electrically connected initially without the need of voltage annealing (see Fig. S4 in Ref. [24]), the AR spectra are dominated by a pronounced ZBCP but with much smaller V-shape background (likely due to much less contribution from a non-superconducting surface layer), and the ZBCP is suppressed at higher B field ~ 2 T (indicating better-quality superconductor layers at the junctions). Therefore, our control experiments caution that a ZBCP disappearing at much lower magnetic field than the bulk critical field and showing a strong non-saturating V-shape background reflects the existence of a degraded superconducting surface layer in non-ideal N-S junctions, and thus cannot be used as the signature for zero-energy Majorana fermions.

In summary, our experimental results can be consistently explained by Andreev reflection process in combination with the effect of the critical current.[27] For the same piece of superconductor microcrystal, a zero-bias conductance peak occurs for N-S junctions with transparent barrier strength (see Fig. 3), but it is absent in N-S junctions with finite barrier strength (see Fig. 2). Therefore, the zero-bias conductance peak observed in $Cu_xBi_2Se_3$ does not represent an evidence for the existence of zero-energy Majorana surface bound states for odd-parity topological superconductors. This draws attention that further investigation is needed to determine if $Cu_xBi_2Se_3$ is indeed the exciting topological superconductor found, important for the identification of potential topological superconductors in future.

Figure Captions

Figure 1. (a) Magnetic susceptibility as a function of temperature for the bulk material of $Cu_xBi_2Se_3$. (b) Optical microscope image of a device with a suspended microcrystal of $Cu_xBi_2Se_3$ on top of multiple parallel metal electrodes (35 nm Pd/ 5 nm Cr) spaced ~ 500 nm apart. The four center electrodes (labeled as 2-5) are designed to be 1 μm wide while the two outside electrodes (labeled as 1 and 6) are 4 μm wide. For this sample, four metal electrodes are electrically connected to the superconductor microcrystal and the corresponding Andreev reflection spectra are shown in Figs. 2 and 3. (c) Schematic diagram of a superconductor microcrystal lying on top of three normal metal electrodes and a circuit designed for obtaining Andreev reflection spectra for the target N-S junction between the $I_+$ ($V_+$) electrode and the superconductor.

Figure 2. Normalized differential conductance $dI/dV$ (symbols) versus bias voltage $V$ at temperature $T = 240$ mK and magnetic field $B = 0$ for three N-S junctions and the fitting to the data (solid line) by the BTK theory (see main text). The fitting parameters are: (a) $\Delta = 0.32$ meV, $\Gamma = 0.21$ meV, $w = 61\%$, and $Z = 0.7$; (b) $\Delta = 0.35$ meV, $\Gamma = 0.13$ meV, $w = 13\%$, and $Z = 0.45$; (c) $\Delta = 0.39$ meV, $\Gamma = 0.27$ meV, $w = 80\%$, and $Z = 0.38$. The data of (a) and (b) are from the N-S junctions at electrode # 2 and #3, respectively, for the sample shown in Fig. 1b, and the data of (c) is from a different sample. The normalized $dI/dV$ is obtained by dividing the measured $dI/dV$ data at T =240 mK by appropriate normal state data at $T$ above bulk $T_c$. For the data of (c), the normal state is not reached

yet within the measured range of the bias voltage (see Fig. S1 of Supplemental Material for the temperature dependence of AR spectra for this N-S junction).

Figure 3. $dI/dV$ vs. $V$ at $T = 240$ mK and $B = 0$ for three N-S junctions with smaller normal state resistance $R_N$. The junction of (a) is from the same sample as the junction of Fig. 2c, and the junctions of (b) and (c) are at electrode #1 and #6, respectively, for the sample of Fig. 1b.

Figure 4. $dI/dV$ vs. $V$ measured under different magnetic fields perpendicular to the substrate at $T = 240$ mK (left columns) and at different temperatures with $B = 0$ (right columns): (a)&(b) for the N-S junction of Fig. 2a, and (c) & (d) for the junction of Fig. 3a.

Figure 5. $dI/dV$ vs. $V$ measured at $T = 240$ mK and $B = 0$ (top row); under different magnetic fields perpendicular to the substrate at $T = 240$ mK (center row); and at different temperatures with $B = 0$ (bottom row) for two N-S junctions of the same superconductor microcrystal in a device. These two N-S junctions were not electrically connected initially, but electrical connection was established after a short voltage pulse (~ 10 V in magnitude) was applied to anneal the contacts between the electrodes and the superconductor microcrystal.

Figures

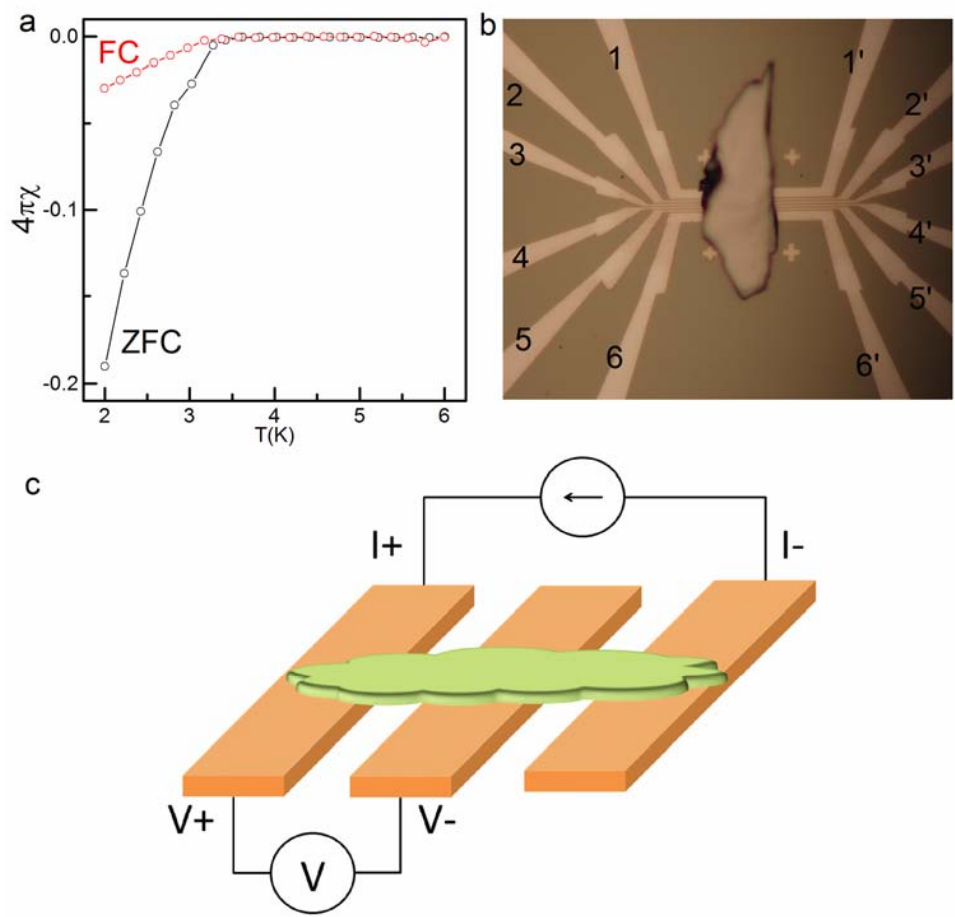

Fig. 1

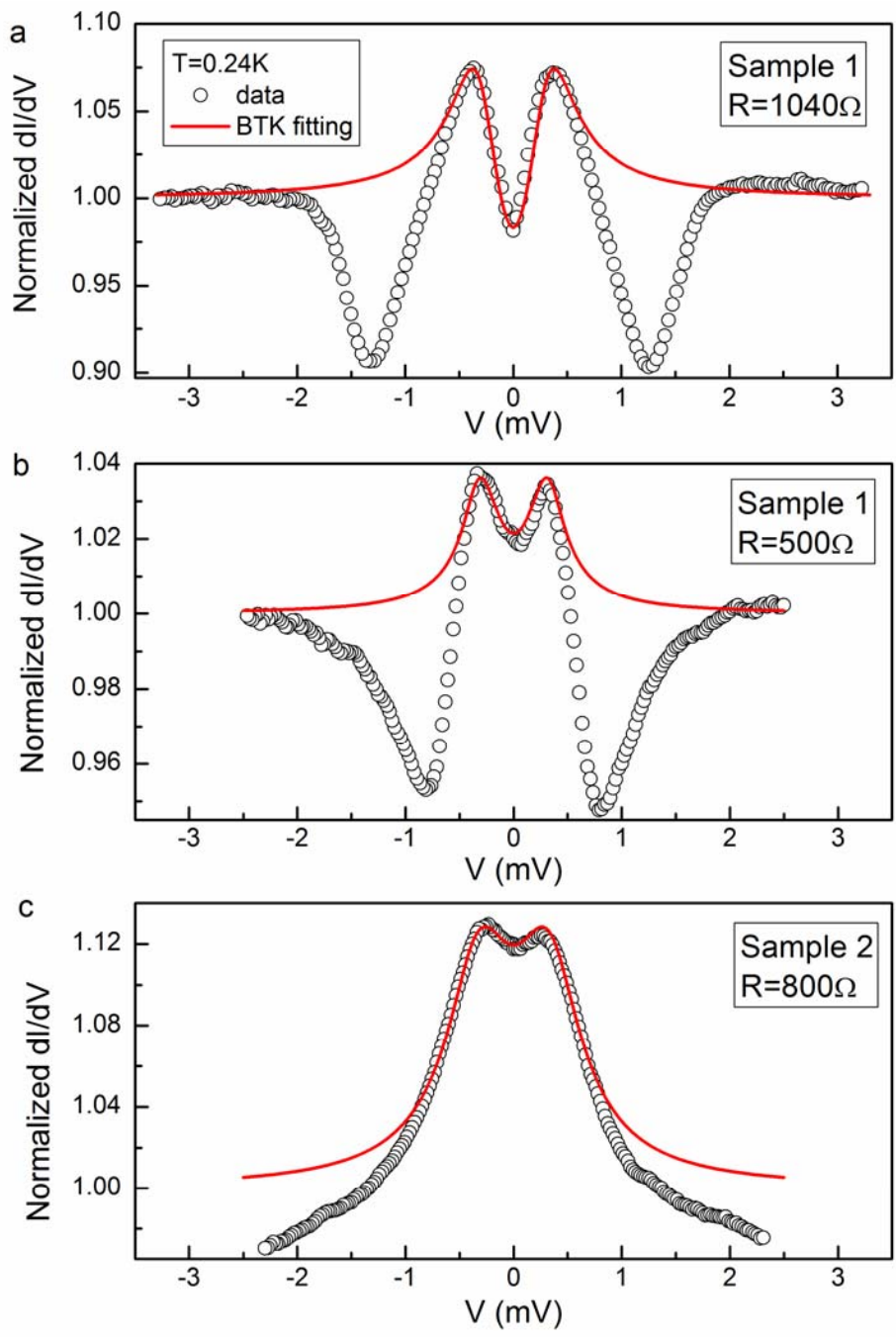

Fig. 2

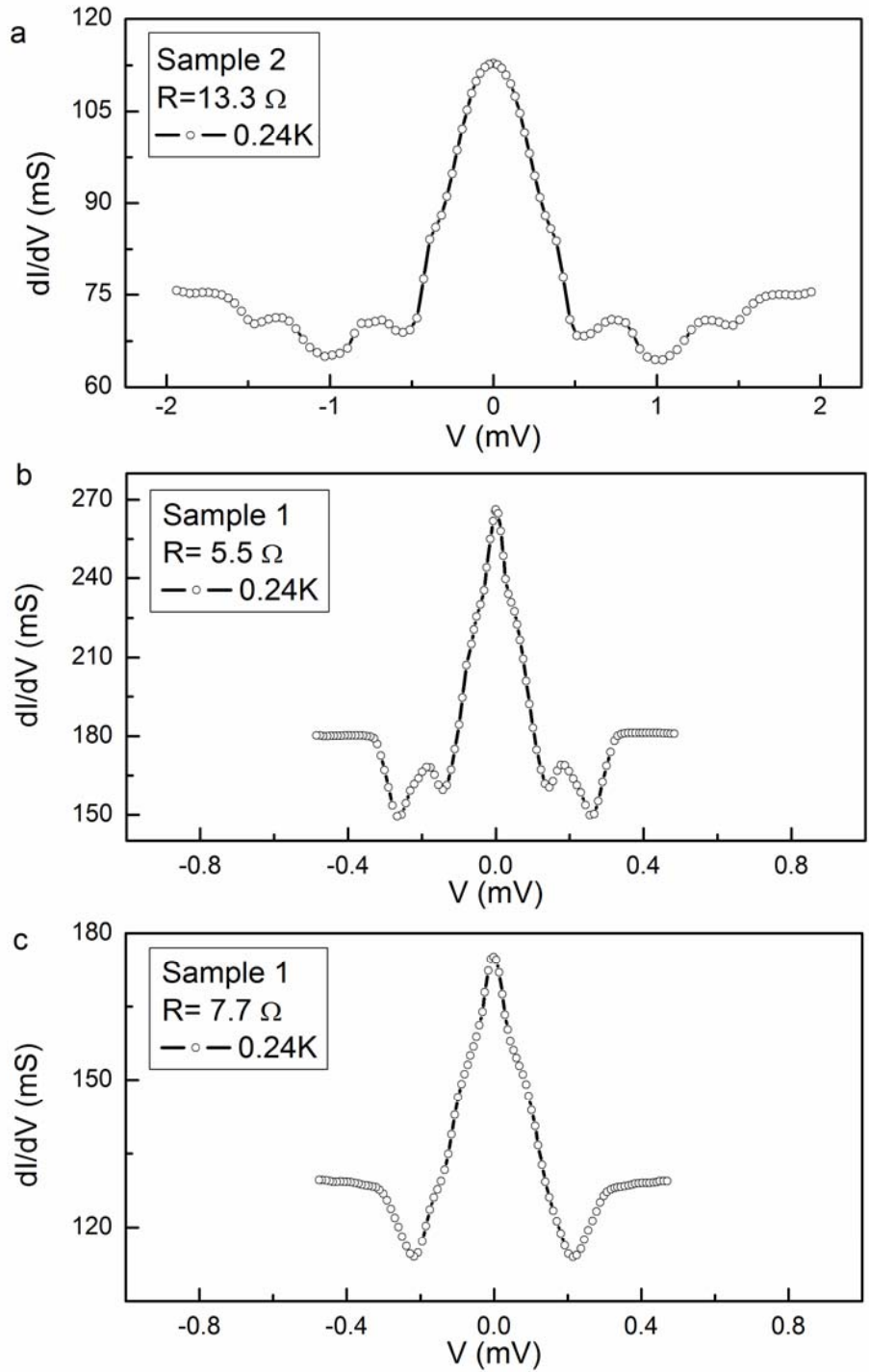

Fig. 3

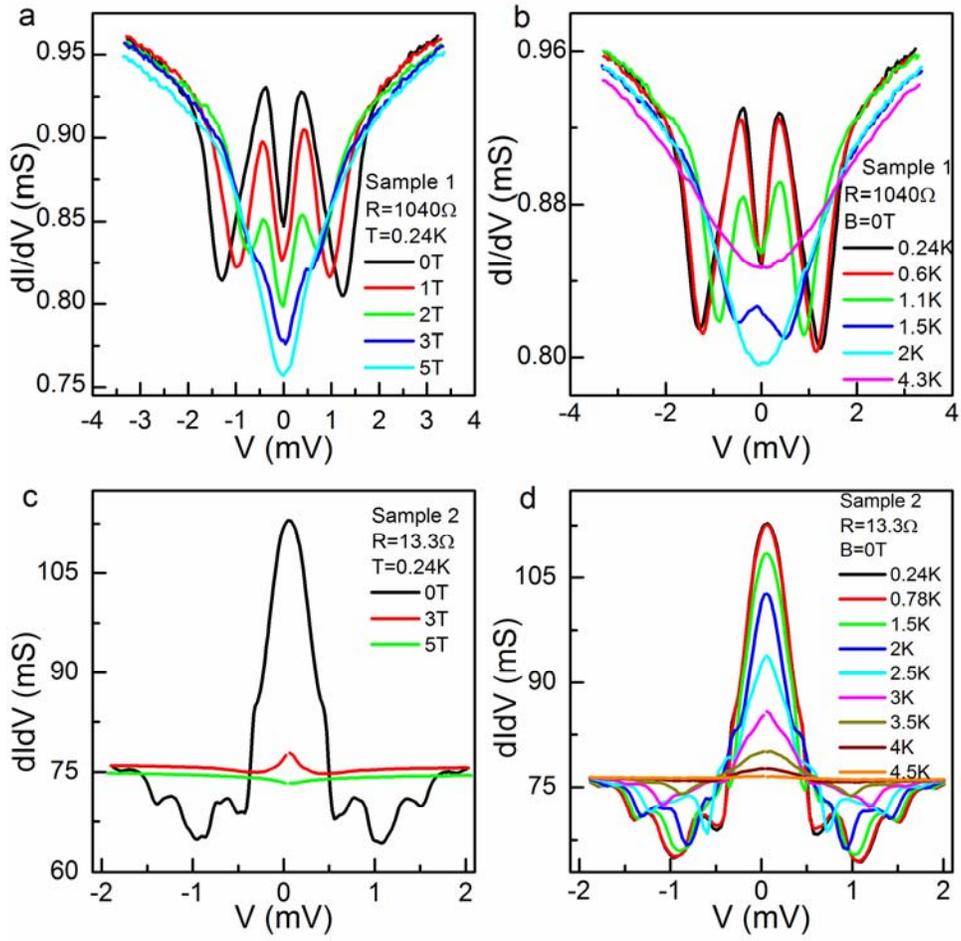

Fig. 4

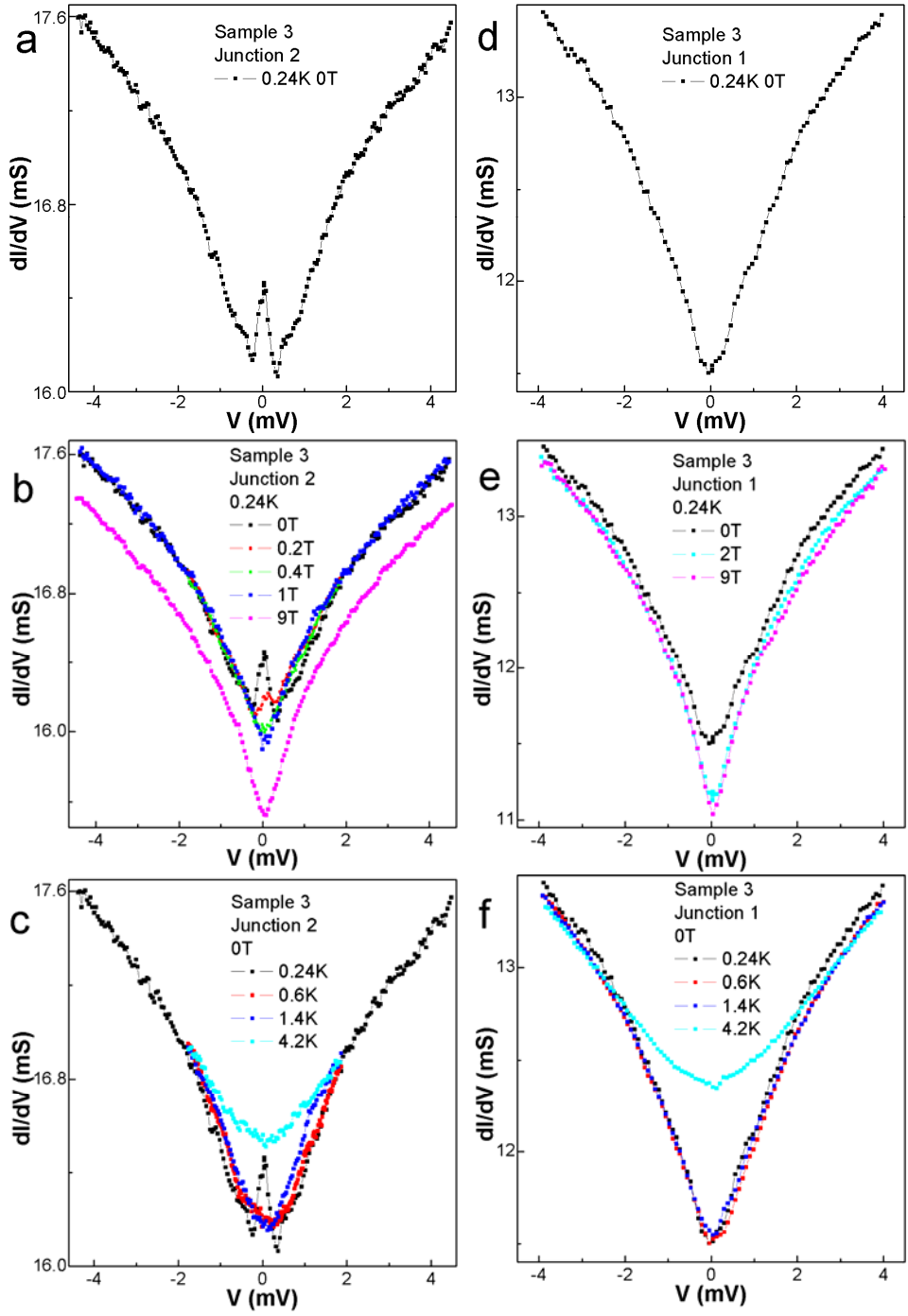

Fig. 5

Supplemental Material for:

# Absence of zero-energy surface bound states in $Cu_xBi_2Se_3$ via a study of Andreev reflection spectroscopy

Haibing Peng, Debtanu De, Bing Lv, Fengyan Wei, Ching-Wu Chu

Department of Physics and the Texas Center for Superconductivity, University of Houston, Houston, Texas 77204-5005, USA

I. A discussion on the determination of transport regimes in a N-S junction to obtain spectroscopic (i.e. energy-resolved) information

In principle, the superconducting gap energy can only be measured accurately in either the ballistic transport regime with the actual N-S point contact radius *a* much less than the electron mean free path *l*, or in the diffusive regime with no significant inelastic scattering (albeit introducing a non-ideal effect of reducing the Andreev refection ratio). In the so-called thermal regime with significant inelastic scattering, the Andreev reflection spectrum can be distorted by energy-changing scattering events and the gap energy may not be obtained accurately from the experimental data. For ballistic conduction through a restriction (e.g., a typical point contact with a radius *a*), the electrical resistance at the normal state can be expressed by the Sharvin formula as $R_N = (4\rho l)/(3\pi a^2)$, with $\rho$ the bulk material resistivity of the normal state for the superconductor. In the case of only one point contact dominating the conduction in a real N-S junction, the ballistic condition $a \ll l$ can then be turned into a condition on the normal-state contact resistance: $R_N \gg 4\rho/3\pi l$. This has been commonly used as an empirical criterion for determining ballisticity, since the normal-state resistance $R_N$ can be measured in experiments (but the radius of a real point contact is experimentally inaccessible in general). More information on the use of the normal-state resistance to estimate the point-contact size and thus determine the ballisticity of a junction can be found in review papers on this topic, e.g., pages 3158-3159 in *J. Phys.: Condens. Matter* **1,** 3157 (1989) (by Duif et al.); page 8 in *arXiv:physics/0312016v1* ( by Naidyuk and Yanson); and pages 2-4 in S*upercond. Sci. & Tech.* **23,** 043001 (2010) (by Daghero and Gonnelli).

In the case of multiple parallel point contacts contributing to the conduction in a N-S junction, the normal-state resistance of each individual contact is larger than the measured total normal-state resistance $R_N$ (i.e. the resistance of the whole N-S junction). Therefore, as long as the measured normal-state resistance $R_N \gg 4\rho/3\pi l$, the N-S junction (either with single or multiple contacts) showing Andreev reflection should be ballistic.

## II. Supplementary figures (Fig. S1 - S4)

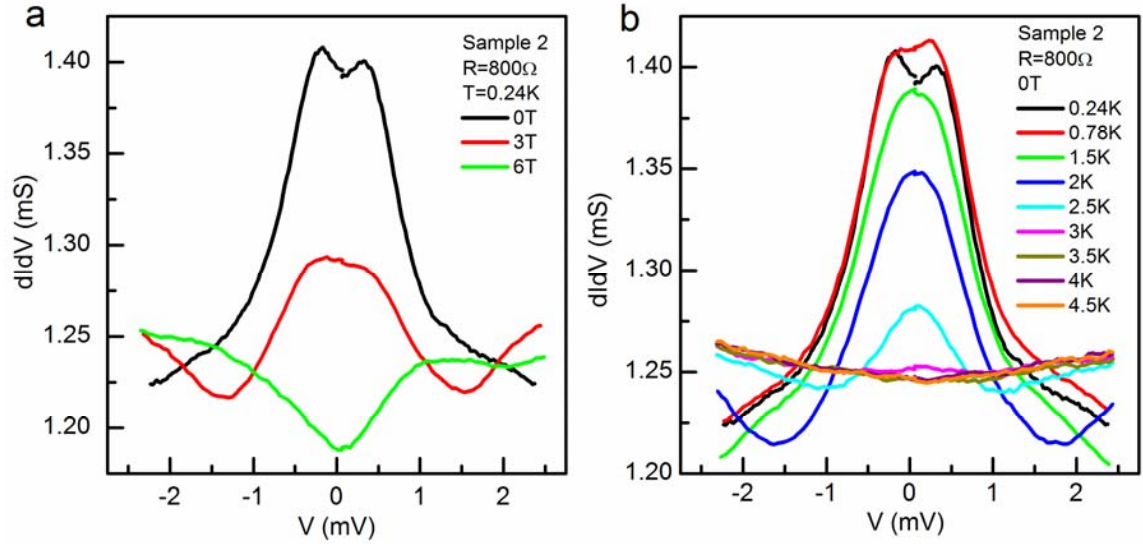

Fig. S1. *dI/dV* vs. *V* measured (a) under different magnetic fields perpendicular to the substrate at $T$ = 240 mK and (b) at different temperatures with $B$ = 0 for the N-S junction shown in Fig. 2c of the main text.

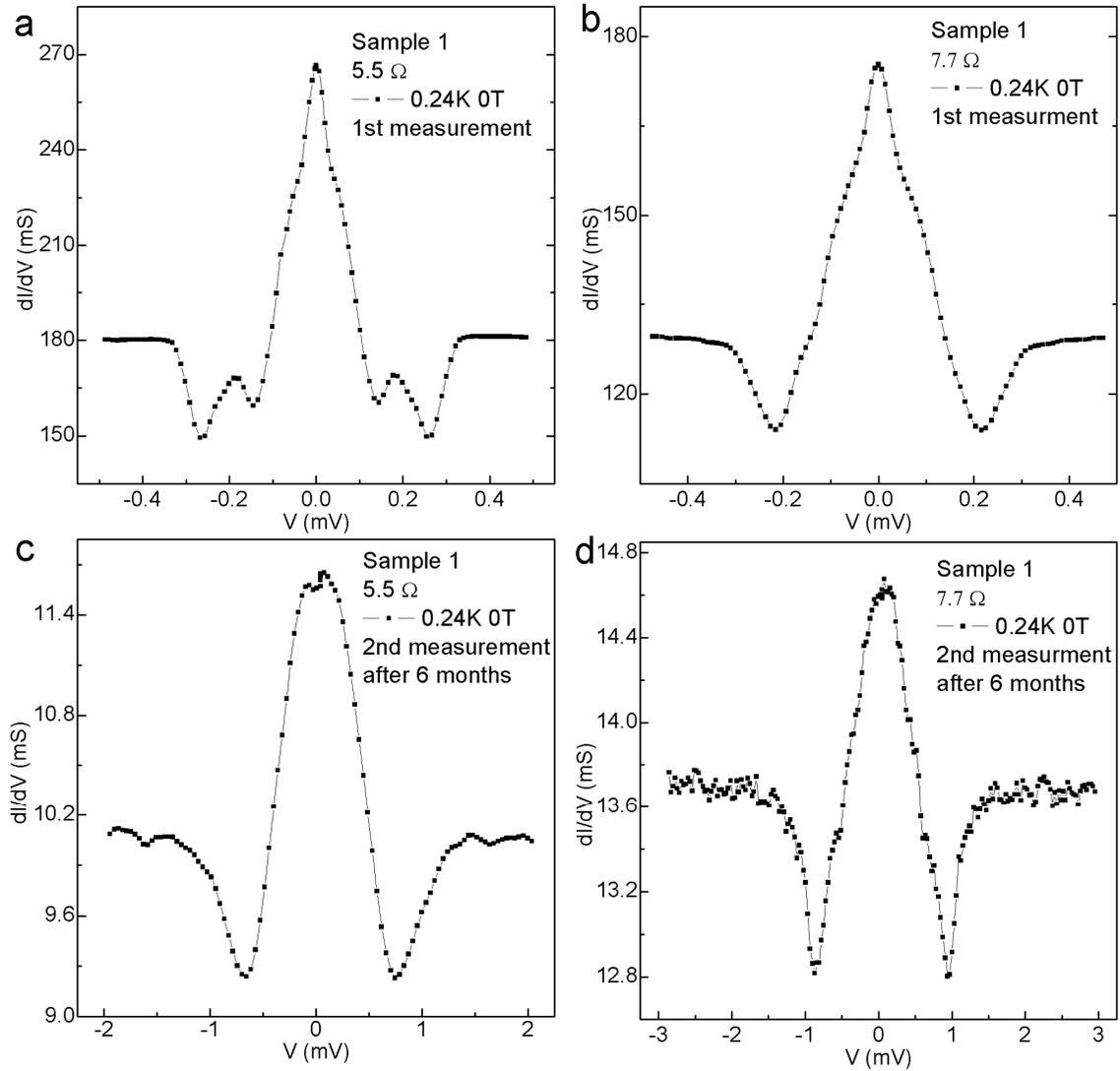

Fig. S2 Comparison between AR spectra in the first measurement (top, replots of Figs. 3b and 3c in the main text) and in the second measurement after a period of six months for the same junctions shown in Figs. 3b and 3c of the main text. The device was preserved in a desiccator during the period between the measurements. The image of the device is shown in Fig. 1b of the main text. In the second measurement, only three electrodes (labeled as #1, #2, and #6 in Fig. 1b) are still electrically connected to the superconductor crystal, which are usable for implementing the measurement by the circuit of Fig. 1c in the main text. The junction shown in the left (right) column is at electrode #1 (#6).

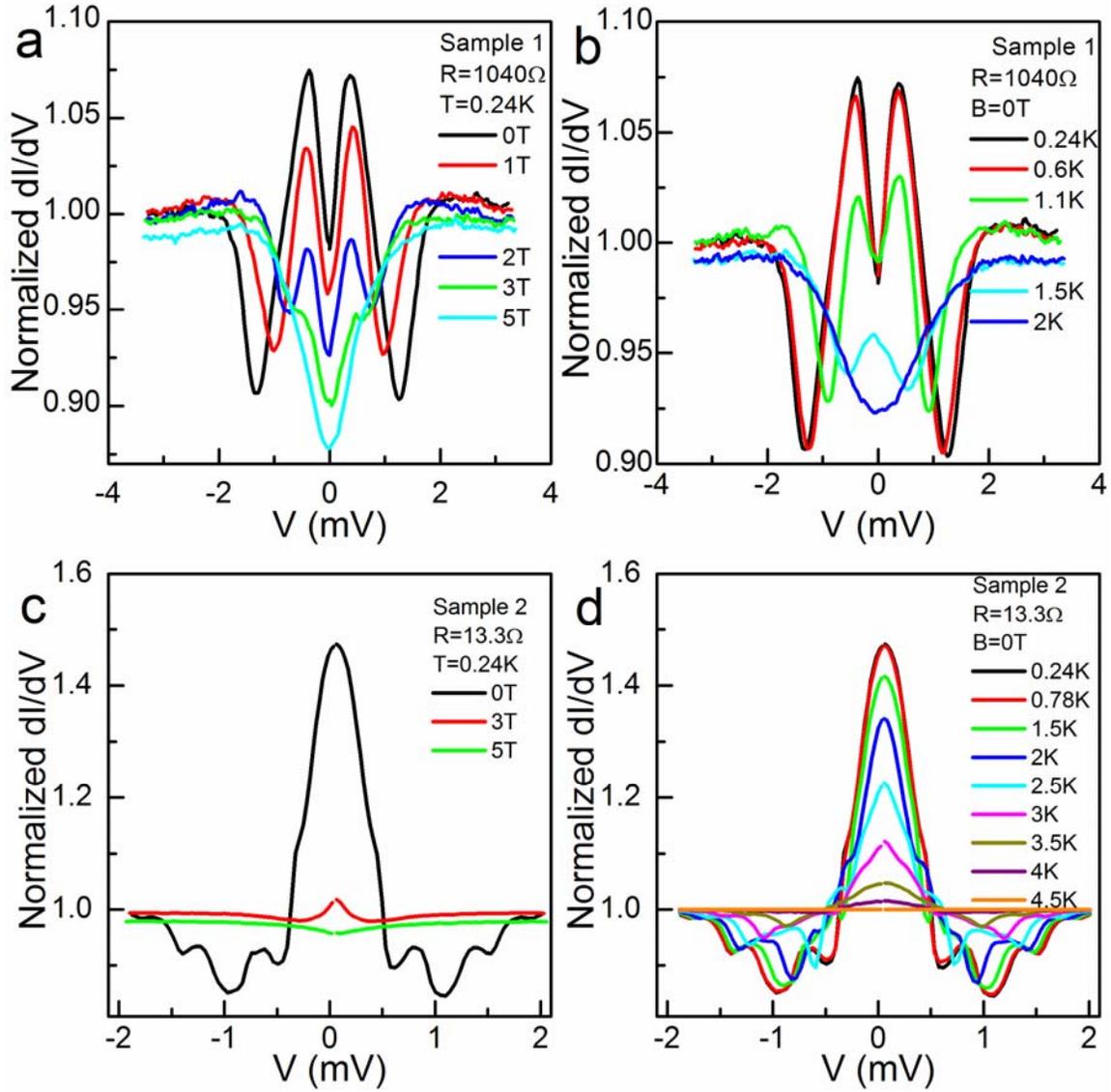

Fig. S3 The normalized dI/dV data for the junctions shown in Fig. 4 of the main text. For (a) and (b), the raw dI/dV data are divided by adjusted normal-state data which are obtained by offset the raw dI/dV spectrum at T = 4.3 K to match the high-bias dI/dV spectrum at T = 240 mK. For (c) and (d), the raw dI/dV data are divided directly by the raw dI/dV spectrum at T = 4.5 K.

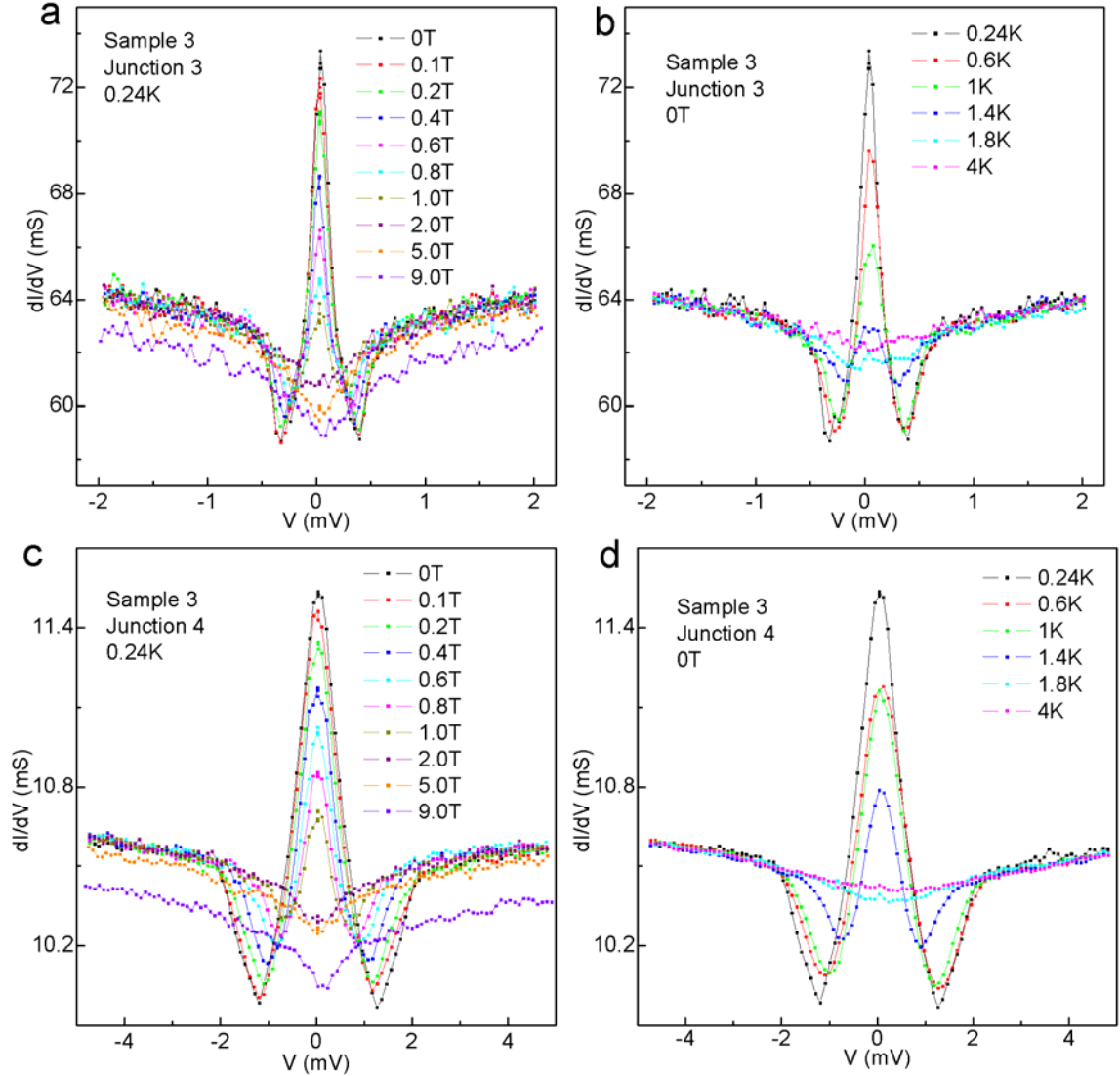

Fig. S4  $dI/dV$ vs. $V$ measured under different magnetic fields perpendicular to the substrate at $T$ = 240 mK (left column) and at different temperatures with $B$ = 0 (right column) for other two junctions in the same device of Fig. 5 in the main text. These two junctions were electrically connected initially without the need of voltage annealing, and they are located at the two outermost electrodes (i.e., #1 and #6) similar to the pattern shown in Fig. 1b of the main text. With four electrically connected junctions in this device, we also obtained the four-terminal resistance of the superconductor microcrystal to be 0.2 Ω above $T_c$ (typical for other devices), which is negligible compared with the junction normal-state resistance $R_N$.